\documentclass[twocolumn,showpacs,preprintnumbers,amsmath,amssymb]{revtex4}

\usepackage{graphicx}

\begin{document}

\title
{Spin and Valley dependent analysis of the two-dimensional low-density
electron system in
Si-MOSFETS.
 }

\author
{
 M.W.C. Dharma-wardana\cite{byline1}
and Fran\c{c}ois Perrot$^*$
}
\affiliation{
Institute of Microstructural Sciences,
National Research Council of Canada, Ottawa, Canada. K1A 0R6\\
}
\date{\today}
\begin{abstract}
The 2-D electron system (2DES) in Si metal-oxide field-effect transistors
(MOSFETS) consists of two distinct electron fluids interacting
with each other. We calculate the total energy as a function of the density
 $n$, and the spin
polarization $\zeta$ in the 
strongly-correlated low-density
regime, using a classical mapping to
a hypernetted-chain (CHNC) equation inclusive of bridge terms.
 Here the ten distribution functions, 
arising from spin and valley indices, are self-consistently calculated to obtain
the total free energy, the chemical potential, the
compressibility and the spin susceptibility.
 The $T=0$ results are compared with the 2-valley Quantum
Monte Carlo (QMC) data of Conti et al. (at $T=0$, $\zeta=0$) and found to be
in excellent agreement.  However, unlike in the
one-valley 2DES, it is shown that
{\it the unpolarized phase is always the stable phase in the 2-valley system},
 right up to Wigner Crystallization
at $r_s=42$. This leads to the insensitivity of
$g^*$ to the spin polarization and to the density.
 The compressibility and the spin-susceptibility enhancement
calculated from the
free energy confirm the validity of a simple approach to the
 two-valley response
based on coupled-mode formation.
This enables the use of
the usual (single-valley) exchange-correlation
functionals in quantum calculations of MOSFET properties
provided mode-coupling effects are taken into account.
The enhancement of the spin susceptibilty calculated
from the coupled-valley response and directly from the
2-valley energies is discussed. The three methods, QMC, CHNC, and 
Coupled-mode theory agree closely. Our results
contain no {\it ad hoc}  fit parameters. They agree with
experiments and do not invoke impurity effects or metal-insulator
transition phenomenology.
\end{abstract}
%
\maketitle
\section{Introduction.}
The 2-D electron systems in GaAs-based structures
as well as those found in Si MOSFETs are being 
intensely studied owing to the accessibility of a
wide range of electron densities under controlled conditions,
leading to a wealth of the experimental observations\cite{krav}.
The nature of the physics depends very much on the
  ``coupling parameter''
$\Gamma$ = (potential energy)/(kinetic energy).
The $\Gamma$ for the 2DES at $T=0$, and the mean density $n$ is
equal to the mean-disk radius $r_s=(\pi n)^{-1/2}$ per electron,
usually expressed in effective atomic units which
depend on the bandstructure mass and ``background''
dielectric constant.
The 2DES in GaAs/AlAs structures  will be
 called a simple 2DES or
 one-valley 2DES
to distinguish it from the two-valley system found in
Si-MOSFETS. 
The inversion layer adjacent to an
oxide layer grown on the Si-(001) surface contains two equivalent valleys
which host the two valleys of the electron fluid.
Various aspects of such multi-valley systems were studied\cite{afs} by
Sham and Nakayama, Rasolt et al, and others, mainly in the high density limit.
The simple 2DES is also a two component
system because of the two spin species, while the 2-valley
system involves 4-components and ten pair-distribution functions (PDFs).

In a recent study of the effective mass $m^*$ and the effective
Land\'e-$g$ factor of the 2DES, the Coulomb coupling between
the electrons of the two valleys was shown to
have a dramatic effect at
low densities when the coupling becomes large\cite{cmm}. In effect, the 
elementary excitations of the two fluids in the two valleys 
interact to form coupled modes,
giving rise to  new effects.
It was shown experimentally\cite{sash} that $m^*g^*$ rises
rapidly with decreasing density in Si-MOSFETS, and that
this rise is due to a dramatic increase in $m^*$, independent of the
spin polarization, while $g^*$
remains essentially constant. Calculations for the
Si system which account for the inter-valley Coulomb coupling
quantitatively predict\cite{cmm} the sharp increase in
$m^*g^*$. 
It was also shown that
$g^*$ remained essentially constant, in strong contrast to the behaviour
found for the simple one valley 2DES\cite{cmm}. The effective mass was also shown to
be practically independent of the spin polarization $\zeta$, 
in excellent quantitative agreement with the data of
of Shashkin et al.\cite{sash} for Si-MOSFETS. This
result leads to the view that the rapid
rise in $m^*$ near $n=1\times10^{11}$ e/cm$^2$ is {\it not caused}
by localization effects, impurity effects, etc., usually
associated with the phenomenology of the  2-D metal-insulator transition.
The enhancement of  $m^*g^*$ in the 1-valley 2DES of GaAs/AlAs systems
is found to be strongly dependent on the spin polarization\cite{jun},
 in strong
contrast to the Si-MOSFET case. Our calculations\cite{cmm} show that
the physics of the 1-valley system is dominated by the presence of a
transition to a fully polarized state which makes $g^*$ increase
rapidly with $r_s$ as the transition density is approached.
 The 2-valley system shows {\it no such
transition to the spin-polarized state} and is insensitive to the
spin polarization.

Given that perturbative many-body theory becomes questionable
for $r_s>1$, within the present context
there are two approaches to evaluating the susceptibility 
enhancement. The second derivative of the total
 free energy $F(n,\zeta,T)$
with respect to the spin-polarization $\zeta$ gives a direct
value for $m^*g^*$ of the two-valley system. This requires the
$\zeta$-dependent 2-valley energy which is not yet available from
quantum Monte Carlo (QMC) simulations.
 However, we can evaluate $F(n,\zeta,T)$
using CHNC, and also show that the CHNC results agree with QMC 
data (available at $\zeta=0$ and $T=0)$.  Another approach,
which avoids the need for a full 4-component calculation is to
build up the 2-valley susceptibility
by noting that the one-valley energy $F(n,\zeta,T)$ is 
available at $T=0$
from QMC, and at any $T\ne 0$ from CHNC. 
The coupling of the excitations of the
two electron fluids in the two valleys can be included in the 
coupled response function in a standard way.
Then we find that the increase in $g^*$ in the GaAs 2DES
 is associated with the "blow-up"
of the single-valley spin response, while the behaviour of the
Si-MOSFET 2DES is related to the "blow-up" of the coupled-mode
response. The static small-$q$ limit of the  
spin response function provides the needed $\chi_s/\chi_p$.

The objective of this paper is: (i) Present $F(n,\zeta)$ data
for the two valley system by a full CHNC calculation involving the
10 pair distributions that are needed in the 2-valley system and
establish the close agreement of the CHNC results
with the QMC calculations.
(ii) Construct the coupled-mode response functions using the well-established
 one-valley
data and show that these results are also validated by QMC and full CHNC 
results.
That is, we present the CHNC results and compare with QMC data
 where available, and establish close agreement. We will not present detailed
finite-T calculations (and hence $m^*$ calculations, as detailed in
ref.~\cite{cmm}) in this paper,  partly to limit the phase space
that has to be studied,
 and since finite-T 
 data are not presently available from QMC for comparison.

In section~\ref{resp} we discuss the construction of the coupled-mode
response functions which use {\it only the one-valley exchange-correlation data}
to obtain the 2-valley behaviour via a physically motivated approximation
to the inter-valley correlations. The compressibility predicted via the
small-$q$ limit of the so constructed 
coupled-mode response is found to agree very
well with that from QMC or the full 4-component CHNC energy calculation.
This validates the coupled-mode model 
used in the calculation (ref.\cite{cmm}) of the $m^*g^*$ enhancement in
Si-MOSFETS. The full 2-valley energy calculations enable us
to examine the 
usual 1-valley local-density approximation (LDA)
 in Si-MOSFETS and the corrections
arising from coupled-mode effects. Finally we discuss the
spin susceptibility enhancement obtained from these calculations, and
the question of relating the electron-disk radius $r_s$ used in
these calculations to the experimental densities.

\section{CHNC calculations for the 2-valley electron fluid.}
\label{chnc}
We consider a 2-DES in a Si-MOSFET at a total density $n$,
with $r_s^2=1/\pi n$,
while the density in each valley $v=a$ or $b$, 
is taken to be $n_v=n/2$. Hence the 
$r_s$ parameter in each valley becomes 
 $r_{sv}=r_s\surd2$. Thus we do not consider density polrizations
leading to $n_a\ne n_b$. Also the electrons in both valleys  have the
same spin polarization $\zeta$ and the same temperature $T$.
This is consistent with recent studies which show that the
valley splitting is very slight\cite{valpud}.
 If the two spin species
be denoted by $i=1,2$, we have a four-component 2-DES with 10 
independent pair distribution
functions (PDFs), viz., $g_{ij,vw}(r)$. We define $k=1,2$ for
 the two spins in valley $a$,
and $k=3,4$ for the two spins in valley $b$, and write the PDFs as $g_{kl}(r)$.
The CHNC method for 2-DES has been described fully in
ref.~\cite{prl2}, where the quantum fluid at $T=0$ is considered to be equivalent to a
classical fluid at a quantum temperature $T_q(r_s)$.
Here, for the two valley system we use the same $T_q(r_s)$ as before,
and embodies the essential ``many-body'' input to the problem.
In brief outline, in CHNC we assume that the 2-D electrons are
mapped on to a classical system where the distribution functions are given by
a finite-$T$ classical density functional form:
\begin{equation}
\label{chnceq}
g_{kl}(r)=e^{-\beta\{P(r)\delta_{ij}\delta_{vw}+ V_{cou}(r) + V_c(r:[g_{kl}])\}}
\end{equation}
Here $\beta P(r)$ is a "Pauli exclusion potential" which acts only for parallel
spins, i.e, if $k=l$. It is
constructed such that $g_{kl}(r)$ becomes identical with the non-interacting
PDF, viz., $g^0_{kl}(r)$ which is known from quantum mechanics
 when the Coulomb interaction
$ V_{cou}(r)$ and the associated correlation corrections $V_c(r)$ are zero. The Coulomb
interaction between two electrons in the equivalent classical picture
 involves a correction arising
from their mutual diffraction effects.
 Thus $ V_{cou}(r)$ is obtained by solving a two-electron
Schrodinger equation. The result is parametrized by the form\cite{prl2},
\begin{eqnarray}
\label{vcou}
V_{cou}(r)&=&(1/r)[1-exp(-k_{th}r)] \\
k_{th}/k_{dBr}&=&1.1587T_{cf}^{0.103} \\
k_{dBr}&=&(2\pi m^*T_{cf})^{1/2}, \; T^2_{cf}=(T^2_q+T^2)
\end{eqnarray}
Here $k_{dBr}$ is the de Broglie momentum of the scattering
pair with the effective pair mass $m^*=1/2$, and $T_{cf}$ is the
classical fluid temperature which reduces to $T_q$ at $T=0$. The correlation
potential $V_c(r)$ occurring in Eq.~\ref{chnceq} is taken to
 be the sum of hyper-netted-chain
diagrams inclusive of a bridge term. Thus $V_c$ is nonlocal and is a
function of the $g_{kl}(r)$ which have to be self-consistently
calculated. The bridge term mimics the higher-order correlations
which are {\it not} captured by the simplest HNC equations.
 These were shown to be important in
 2-D electron systems in reference ~\cite{prl2}. 
Particles having identical indices ($k=j$)
are restricted from close approach by the Pauli
 exclusion effect modeled by $P_{kl}(r)$. However,
singlet-pairs of electrons, or electrons in two different
 valleys contribute to strong Coulomb
correlations, and hence a bridge term is included in all
 such ``off-diagonal'' PDFs. The bridge term
 $B_{kl}(r)$ for $k\ne l$ applies to 6 different PDFs, and we have taken this
to be given by the usual hard-disk functional form discussed in ref.\cite{prl2}.
(Totsuji and coworkers\cite{hard} have studied a more detailed
 implementation of the
hard-disk bridge function in CHNC, while Bulutay et al.\cite{bulut} have
studied the 2-D CHNC without a bridge correction).
It should be emphasized that both the HNC approximation as well as the need for
a bridge function can be avoided by using the classical mapping to
a quantum fluid (CMQF) where we use classical molecular dynamics (MD)
to generate the PDFs etc., of the classical fluid under consideration.
In such a scheme we use the pair-potential given by Eq.~\ref{vcou} plus
the Pauli potential in an MD simulation for a
classical plasma 
at the temperature $T_{cf}$. Such a CMQF-MD scheme would be numerically more
demanding than the CHNC,  much simpler than the full QMC simulations,
 and have the
advantage of not making the HNC+bridge approximations. However, the 2-valley
(4-component) system examined here has been studied by QMC and we use those
results to confirm the validity of our methods.

The main difference in the physics of the 1-valley system, and the
2-valley system arises from the preponderence of direct
Coulomb interactions (from 6 PDFs in the 2-valley, one in the 1-valley)
 over the exchange
interactions (from 4 PDFs in the 2-valley, two PDFs in the 1-valley).
 This is the main reason for the lack
of a transition to a stable $\zeta=1$ state at low density.
Since the transition to a $\zeta=1$ state does not occur as
$r_s$ increases, the $g^*$
remains  insensitive to increasing $r_s$,
as found theoretically (\cite{cmm})
and experimentally\cite{sash}.

	The 10 coupled equations for $g_{kl}(r)$ are
 self-consistently solved for many
values of the coupling constant $\lambda$ applied to
 the Coulomb interaction, and the
$g_{kl}(r:\lambda)$ are used in the adiabatic 
connection formula to determine the 
exchange-correlation free energy of the 2-valley 2DES.
 While our calculations are easily
carried out for any value of $\zeta$, $T$ and $r_s$,
 the 4-component QMC calculations 
at finite $T,\zeta$ are
a major computational  undertaking which has not been attempted.
 However, at $T=0, \zeta=0$,
 Conti and Senatore\cite{cs} have presented QMC results
 for 2-D electron bilayers separated by
a distance $d_L$. They give total energies and also a fit to the
 correlation energy/electron,
$\epsilon_c(r_s,\zeta=0,T=0)$ at $d_L=0$, i.e, the case where
both electron gases reside in the same layer.
 In table~\ref{csdata} we compare the 4-component CHNC with
 the available 4-component QMC data at $d_L=0$.
 The energies $\epsilon_{c}(r_s)^{\scriptscriptstyle QMC}$, 
are from the Rapisarda-Senatore fit-formula\cite{rapi} with
the parameters quoted in Table I of Conti et al\cite{cs}.

\begin{table}
\caption{ Comparison of the total energy $\epsilon_{tot}(r_s)$
and and the correlation energy $\epsilon_{c}(r_s)$, in atomic units
at $T=0,\zeta=0$ for the 2-valley 2DES obtained from CHNC, with the
QMC data of Conti et al\cite{cs}.}
\begin{ruledtabular}
\begin{tabular}{cccccc}
$r_s$ &  $\epsilon_{tot}(r_s)^{\scriptscriptstyle QMC}$&
 $\epsilon_{tot}(r_s)^{\scriptscriptstyle CHNC}$ & 
 $\epsilon_{c}(r_s)^{\scriptscriptstyle QMC}$ & 
 $\epsilon_{c}(r_s)^{\scriptscriptstyle CHNC}$\\
\hline \\
2.0   & -0.29302 & -0.29172   &-0.14315 &-0.14202 \\
10.0  & -0.08611 & -0.08647   &-0.04607 &-0.04649 \\
20.0  & -0.04641 & -0.04643   &-0.02577 &-0.02581 \\
30.0  & -0.03196 & -0.03183   &-0.01806 &-0.01795 \\ 
\label{csdata}
\end{tabular}
\end{ruledtabular}
\end{table}

  These results show that the CHNC method provides a simple and
accurate approach to the treatment of exchange and correlation in
the 4-component system. We exploit the simplicity of CHNC for calculations
at finite-$T$ and $\zeta$ which are currently too expensive for
QMC simulations. However, in situations where QMC results are
available for the correlation energies, we adopt the QMC parametrizations.
Thus one may use the parametrization of $\epsilon_c$ given by Conti and
Senatore for the $T=0, \zeta=0$ case. For $r_s>1$ applications the
Tanatar-Ceperley\cite{tc}  form may also be used for the 2-valley
system as well, with the parameter values
$a_0=-0.40242,\, a_1=1.1319, \, a_2=1.3945, \, a_3=0.67883$, fitted
to a data base from $r_s$ =1 to 30.

  The $T=0,\,\zeta=1$ case is particularly interesting since this system
(i.e, 2-valley system at density $n$) is
mathematically identical to the 1-valley  system at the same density
but with $\zeta=0$, for the Hamiltonian considered by Conti et al.,
and by us in this study. In this case the 2-valley system has a two-fold symmetry
since the  energy is the same irrespective of the orientation of the spin
in each valley. That is, $\zeta=1$ means all the spins in valley $a$ are oriented,
while all the spins in valley $b$ are also oriented, but independently of
the orientation of the spins in $a$. This degeneracy would be resolved in
real Si-MOSFETS but not in the model used here, or in Conti and Senatore.
For instance, the three-body correlations for inter-valley interactions
may be slightly different from those in the {\it intra}-valley interactions,
and hence may require two different bridge parameters, to be determined
variationally by an energy minimization using the hard-disk
reference fluid approach. We have not done this, and simply used
the same bridge parameter as in ref.~\cite{prl2} for all interactions.
In the QMC calculation this would require independent optimization of the
model for back-flow corrections. Finally, 
the correlation energy of the fully spin-polarized (degenerate) 2-valley system 
can be parametrized using the Tanatar-Ceperley form with
$a_0=-0.19162,\, a_1=3.6123, \, a_2=1.9936, \, a_3=1.4714$, in atomic units.
\subsection{The energy of unpolarized and polarized phases.}
   The $T=0$ correlation energy at finite values of $\zeta$ were
 calculated using 
the CHNC procedure and compared with the
values predicted from the  polarization factor used for the
one-valley 2DES. This has the form\cite{prl2}
\begin{eqnarray}
\label{spinpol}
p(r_s,\zeta) &=& \frac{\epsilon_c(r_s,\zeta)-\epsilon_c(r_s,0)}
{\epsilon_c(r_s,1)-\epsilon_c(r_s,0)}
\,\, = \frac{\zeta_+^{\alpha(r_s)}+\zeta_-^{\alpha(r_s)}-2}
{2^{\alpha(r_s)}-2}\nonumber\\
\alpha(r_s)&=&C_1-C_2/r_s+C_3/r_s^{2/3}-C_4/r_s^{1/3}
\end{eqnarray}
Here,  $\zeta_{\pm}$=$(1\pm\zeta)$.
It turns out that the coefficients $C_1-C_4$ obtained
for  the one-valley 2DES, i.e., 
1.5404, 0.030544, 0.29621, and 0.23905 respectively, work quite 
well for the 2-valley system as well,
 even at high $r_s$.
 Thus, using the 
TC-type fit formulae for the 2-valley $\epsilon(r_s,\zeta=1)$ and
$\epsilon_c(r_s,\zeta=0)$, the estimated values of $\epsilon_c(r_s,\zeta)$
and the CHNC values are given 
in Table~\ref{polfit}.
\begin{table}
\caption{ The correlation  energy $\epsilon_{c}(r_s,\zeta)$
per electron, as a function of $\zeta$, estimated using the 1-valley
polarization factor of Eq.~\ref{spinpol}, and from the full 2-valley
CHNC calculation}
\begin{ruledtabular}
\begin{tabular}{cccccc}
$r_s$ &  $\epsilon_{c}^{fit}$& 
$\epsilon_{c}^{\scriptscriptstyle CHNC}$ &
 $\epsilon_{c}^{fit}$ &  $ \epsilon_{c}^{\scriptscriptstyle CHNC}$\\
\hline \\
$\;\;\;\;\zeta \to$&0.25 & 0.25& 0.75 & 0.75 &\\
\hline \\
5.0   & -0.07686  & -0.07757 &-0.06286 & -0.06434\\
10.0  & -0.04518  & -0.04562 &-0.03765 & -0.03831 \\
20.0  & -0.02529  & -0.02535 &-0.02134 & -0.02151 \\
30.0  & -0.01772  & -0.01764 &-0.01477 & -0.01504 \\
\label{polfit}
\end{tabular}
\end{ruledtabular}
\end{table}

The finite-$\zeta$
calculations show that there is  {\it no ferromagnetic phase transition}
in this system at $T=0$, since the total energy of the $\zeta=0$ phase
is always more negative than any polarized phase. This is expected from the
dominance of the many off-diagonal terms contributing direct 
Coulomb interactions, but no exchange interactions.
This was pointed
out in ref.~\cite{cmm} where the insensitivity of $m^*g^*$ to $\zeta$
obtained from the theory was in excellent agreement with the experiments
of Shashkin et al\cite{sash}. The
stabilization energy $\Delta E$ of the $\zeta=0$ phase with respect to
the fully polarized phase is 0.12734$\times 10^{-3}$ a.u. 
at $r_s$=25, and diminishes 
to 0.89929$\times 10^{-4}$ a.u. at
 $r_s=40$. These are very small energy differences
and within the error of the CHNC method and possibly of the 2-valley QMC calculations.
 However, the pattern of 
stability of the $\zeta=0$ phase holds for
 all $r_s$ investigated. Conti and Senatore\cite{cs} find a Wigner crystal
phase at $r_s=42$ and claim that ``in the intermediate regime the 4-component
2-DES can be expected to mimic the 2DES $\cdots$, with the appearance of
a spin-polarized ground state''.  We arrive at the opposite conclusion.
\subsection{LDA-type calculations for Si-MOSFETS.}
\begin{figure}
\includegraphics*[width=9.0cm, height=10.0cm]{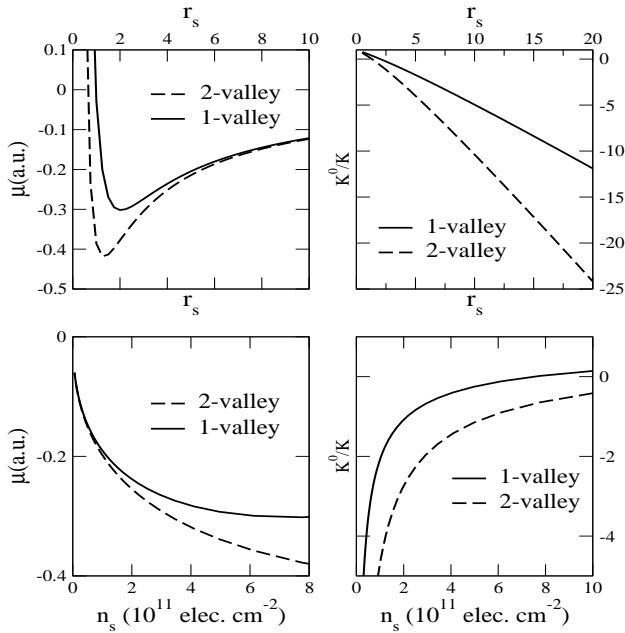}
\caption
{Left panels: Comparison of the total chemical potential $\mu$
i.e, the total Kohn-Sham potential, calculated at
 the total density $n$ and $r_s$, for the 2-valley system,
 and if the LDA were used (labeled 1-valley), ignoring the 2-valley 
nature. Right panels: Same for the compressibility ratio}
\label{muk}
\end{figure}

	In most density-functional calculations of Si/SiO$_2$ 
quantum wells, the
Kohn-Sham exchange-correlation potential $V_{xc}(r)$
 is calculated using the local
density approximation (LDA) where the total density
 $n(r)$ is considered without
taking account of the valley degeneracy. In effect, the electron gas is assumed
to be a {\it single} electron gas at a density $r_s$ and its exchange-correlation
energy $\epsilon_{xc}(r_s)$ and the Kohn-Sham potential $V_{xc}(r_s)$ are calculated
at the given density.
(We recall that $V_{xc}(r_s)$ is simply the xc-contribution to the 
chemical potential $\mu_{xc}$). 
Results for $\epsilon_{xc}$ and $\mu_{xc}$ for electrons in a Si-MOSFET calculated
correctly, i.e., taking account of its degenerate valley structure, and in the usual
LDA approach are compared in Table~\ref{lda}.
The full chemical potential $\mu$, as well as the compressibility ratio $K^0/K$
calculated for the 2-valley system, and that obtained within the LDA approach,
are also shown in Fig.~\ref{muk}. It is noteworthy that
the total chemical potential has a minimum near $r_{sm}\sim 1$. In effect, a low density
 electron fluid ($r_s > r_{sm}$) whose chemical potential is
equal to  that of a high density
gas ($r_s < r_{sm}$) exists and this could lead to spontaneous
density inhomgenieties in these systems.
%
%
\begin{table}
\caption{ Comparison of the exchange-correlation  energy $\epsilon_{xc}(r_s)$
per electron, and the Kohn-Sham potential $V_{xc}(r_s)$, in atomic units
at $T=0,\zeta=0$ for the 2-valley 2DES obtained from QMC-fit or  CHNC, and from
the LDA.}
\begin{ruledtabular}
\begin{tabular}{cccccc}
$r_s$ &  $\epsilon_{xc}(r_s)^{LDA}$& $\epsilon_{xc}(r_s)^{QMC}$ &
 $V_{xc}(r_s)^{LDA}$ &  $ V_{xc}(r_s)^{QMC}$\\
\hline \\
2.0   & -0.38213  & -0.35535 &-0.55189 & -0.50271 \\
10.0  & -0.09007  & -0.08851 &-0.13133 & -0.12835 \\
20.0  & -0.04742  & -0.04699 &-0.06957 & -0.06874 \\
30.0  & -0.03241  & -0.03221 &-0.04769 & -0.04730 \\
\label{lda}
\end{tabular}
\end{ruledtabular}
\end{table}

 When the actual electron densities in Si-MOSFETS are converted to effective $r_s$
units (see below), the $r_s$ range 1-6 is the most important for device
applications, and hence overestimates in $V_{xc}$ contained in the usual LDA
approach could be significant.

\subsection{Pair-distribution functions in the Si-MOSFET system}
\begin{figure}
\includegraphics*[width=9.0cm, height=10.0cm]{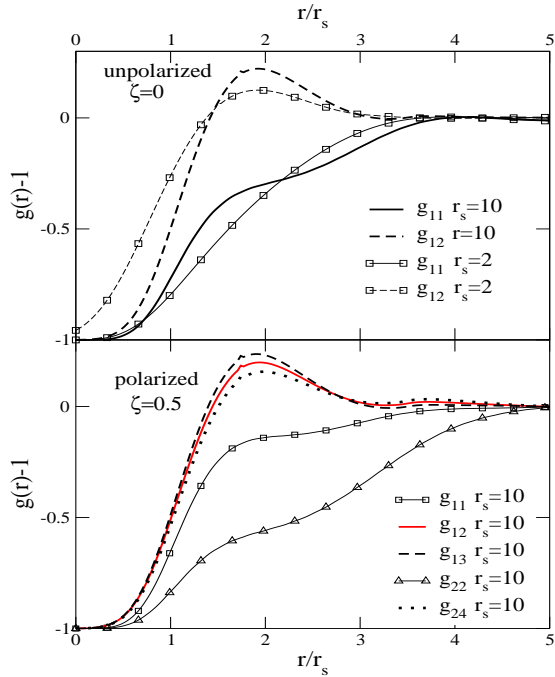}
\caption
{Pair distribution functions of the two-valley system. For $\zeta=0$
 the 10 PDFs reduce to two indeependent PDFs. For $\zeta \ne. 0$
 there are 5 independent PDFs.
}
\label{pdf}
\end{figure}

	The PDFs, denoted by $g_{kl}(r)$, embody the detailed
particle correlations in
the system. In Fig.~\ref{pdf} we display an illustrative set of 
pair-distribution
functions. QMC-based
PDFs have not been reported in the literature and hence we do not have a
direct comparison. However, good agreement between CHNC and QMC-
based PDFs have been found in other systems (e.g, 2DES, 3DES, and fluid
hydrogen)\cite{prl2,prl1,hyd}.
For $\zeta=0$ the four
diagonal PDFs  $g_{kk}$ are
 identical, and similarly, all the six
off-diagonal PDFs are also identical.
 Hence there are actually only two distinct PDFs, just
as in the single-valley case where $g_{11}$ and $g_{12}$ define the $\zeta=0$
case. These are shown for $r_s=2$, 5 and 10 in the top
panel of Fig.~\ref{pdf}.
 It is also clear that the correlation effects are mainly
determined by the off-diagonal PDFs. At finite $\zeta$ there are five
independent PDFs. There are two independent diagonal PDFs 
$g_{11}=g_{33}$ and $g_{22}=g_{44}$. The three independent off-diagonals
are $g_{12}=g_{23}=g_{34}=g_{14}$, $g_{13}$ and $g_{24}$.
These are shown for the case $r_s=10$ and $\zeta=0.5$ in the
lower panel of Fig.~\ref{pdf}. 
\section{Response functions.} 
\label{resp}
 In the following we  discuss the linear response functions
 since the static small-$k$ limit can be related to the
 derivatives of the total energies that were calculated from CHNC or QMC
 if available.
 This enables us to verify a simple procedure for the construction
 of the 2-valley response functions using {\it only} single-valley
 exchange correlation data\cite{cmm}. The density-density response
function $\chi(k,\omega)$ will to be called the  {\it d-d} response for brevity.
It is  expressed
in terms of a reference ``zeroth-order''
 $\chi^0_R(k,\omega)$ and a local-field factor (LFF), 
denoted by $G(k,\omega)$.
\begin{equation}
\label{lffdef}
\chi(k,\omega)=\chi^0_R(k,\omega)/[1-V_k\{1-G_d(k,\omega)\}\chi^0_R(k,\omega)]
\end{equation}
The {\it s-s} response (or ``spin susceptibility'') is written as
\begin{equation}
\label{spindef}
\chi_s(k,\omega)=-\mu_B^2\chi^0_R(k,\omega)/[1-V_k\{1-G_s(k,\omega)\}\chi^0_R(k,\omega)]
\end{equation}
where $\mu_B$ is the Bohr magneton. Note that our definition of
the spin-LFF differs somewhat from a commonly used
definition\cite{santoro}.
 Our form makes the {\it d-d} and
{\it s-s} LFFs appear formally similar, at least at this stage.
Hence a single discussion applies
to both, and we drop the subscripts $d$ and $s$.
The reference function $\chi^0_R(k,\omega)$ has the form:
\begin{equation}
\chi^0_R(k,\omega)=\sum_{\vec{k},\sigma} \frac{n_{\sigma,\vec{k}}
-n_{\sigma,\vec{k}+\vec{q}}}
{\omega+\epsilon_{\vec{k}}-\epsilon_{\vec{k}+\vec{q}}}
\end{equation}
Here $\vec{k},\vec{q}$ are two-dimensional vectors, while the corresponding
single-particle energies are denoted by $\epsilon_{\vec{k}}$ etc. The 
Fermi occupation number $n_{\sigma,\vec{k}}$ may be chosen to be the
non-interacting value, in which case $\chi^0$ is the 2-D Lindhard
 function\cite{afs}. An alternative choice is to use the
fully-interacting density, evaluated from the fully-interacting
chemical potential, as
in density-functional theory (DFT).

 The small-$k$ limits of the local-field factors $G_d(k,\omega)$ and $G_s(k,\omega)$
can be obtained
from the second derivatives
of the exchange-correlation free-energy functional $F_{xc}(n,\zeta)$ of DFT,
 with respect to
the charge densities or the spin densities\cite{iwamoto}.
 These second-order derivatives, together
with the second derivative of $F_{xc}(n,\zeta)$ with respect to $T$
for a two-valley system were used in our $m^*$ and $g^*$ calculations
 reported in ref.\cite{cmm}, using the CHNC technique.
 Here we look at the compressibility and spin-susceptibility
 of the 2-valley system obtained from
 the small-$k$ limit of the coupled-mode response function (built up
 from 1-valley data) and compare it
 with that obtained directly from the 2-valley CHNC and QMC.
\subsection{Response functions of the two-valley system.}
%
    The theory of the one-valley fluid can be used for the two-valley
(4-component) fluid if there
is no valley polarization (i.e, the two valleys are assumed equivalent
although distinct), as in
ref.\cite{cmm}. As this may not be completely
clear from the abbreviated discussion in ref.\cite{cmm}, we 
present some details here.

In the theory of classical fluids, the response functions are
simply related to the structure factors, while the LFFs are
simply related to the direct correlation functions
of Ornstein-Zernike(OZ) theory. Since this paper is directed
more towards electron-fluid studies, we follow the language
of the LFFS and the related response methodology rather than the
OZ presentation.

Let us indicate the species 
(which may be a valley index or a spin index) 
 by $u$ or $v$, taking the values 1,2, and let
us consider a weak external potential  $\phi_v(\vec{k},\omega)$ which
acts only on the electrons of species $v$. The external potential induces
density deviations $\delta n_v(\vec{k},\omega)$ such that\cite{vashista}:
\begin{equation}
\label{linres}
\delta n_u(\vec{k},\omega)=\sum_v 
\chi_{uv}(\vec{k},\omega)i\phi_v(\vec{k},\omega)
\end{equation}
These equations define the linear {\it d-d}
 response functions involving the
species $u$ and $v$. The longitudinal
dielectric function $\varepsilon(\vec{k},\omega)$ is now given by
\begin{equation}
1/\varepsilon(\vec{k},\omega)=1+V_k\sum_{u,v}\chi_{uv}(\vec{k},\omega)
\end{equation}
Here $V_k$ is the 2-D Coulomb interaction $2\pi/k$. Note that we are using
effective atomic units (Hartrees etc.) such that 
 $e^2$ divided by the background dielectric constant
is taken to be unity (see section~\ref{n1n2}).
 To relate the response functions to the local fields,
we consider the effective potentials $U_v(\vec{k},\omega)$ such that:
\begin{equation}
U_v(\vec{k},\omega)=V_k[1-G_{uv}(\vec{k},\omega)]\delta n_v(\vec{k},\omega)
\end{equation}
Thus the bare Coulomb interaction between the particles of type $u$ and $v$
is
modified by the LFFs $G_{uv}$. Hence we can write the density deviations
$\delta n_u(\vec{k},\omega)$ in terms of the effective potentials and the
zeroth order response functions as follows (we drop the $\vec{k},\omega$ labels for brevity).
\begin{equation}
\label{effective}
\delta n_u=\chi^0_u[\phi_u
+V_k\sum_v (1-G_{uv})\delta n_v]
\end{equation}
Now, by a comparison of the equations \ref{linres} and\ref{effective}, we can
write down the response functions of the coupled  two-component system in terms
of the zeroth-order response functions and the LFFs.
\begin{eqnarray}
\label{coupled}
\chi_{11}&=&\chi_1^0\,d_2/D,\,\,\chi_{22}=\chi_2^0\,d_1/D, \\
\chi_{12}&=&V_k\chi_1^0\chi_2^0[1-G_{12}]/D \\
d_1&=&1-V_k\chi_1^0[1-G_{11}]\\
d_{12}&=&V_k\chi_1^0[1-G_{12}]\\
 D&=&d_1 d_2 -d_{12}d_{21}
\end{eqnarray}
We have suppressed the $\vec{k},\omega$ dependence in the above
 equations, and also not explicitly given $\chi_{21}$, $d_2$ and $d_{21}$  for brevity.
We now define a total coupled-mode response function $\chi_T(\vec{k},\omega)$ via
\begin{equation}
1/\varepsilon(\vec{k},\omega)=1+V_k\chi_T(\vec{k},\omega)
\end{equation}
Then the total two-component 2DES response is given by:
\begin{eqnarray}
\label{sumichi}
\chi_T&=&[\chi_1^0+\chi_2^0+V_k\chi_1^0\chi_2^0G_{\Sigma}]/D\\
G_{\Sigma}&=&G_{11}+G_{22}-G_{12}-G_{21}
\end{eqnarray}
In the following we some times denote $\chi_T$ by $\chi_{cm}$ to
emphasize the coupled-mode nature of the total response.
If we are dealing with a simple (one-valley) electron fluid,
e.g., a partially spin-polarized electron gas,
the species index $v$ is simply the
spin index.
Notice that the coupling between the two systems (be they spins or valleys),
replaces the individual denominators $d_1$ and $d_2$ by a new denominator
$D$, common to both systems, and containing the  cross terms $d_{ij}$.
 That is, instead of the two sets of excitations given by the zeros
of $d_1$ and $d_2$, we now have a {\it common set
 of ``coupled-mode'' excitations}
defined by the zeros of $D$.
We emphasize that in this analysis we have {\it not }
used any form of CHNC theory.

 All the response functions prior
to the switching-on of the Coulomb interaction between the
two valleys are known. The problem is to determine the
cross terms $d_{ij}$, i.e, the inter-valley term, $d_{uv}$, occurring
in the coupled-mode denominator $D$, using only the free energy data
for a single valley.
If we consider the
small-$k$ limit, we see that the terms $d_{ij}$ occurring in
$D$ are directly related to the second  density derivative or magnetization
derivative of the free energy contributions $F_{ij}$ arising
from the PDFs $g_{ij}$.  We know these {\em individual}
free energy contributions  for the one-valley problem.

Let us first consider the small-$k$
limit of the static response functions to make contact
with the compressibility and susceptibility sum rules.

\subsection{Small-$k$ limit of the static response.}
\label{smallk}
The small-$k$ behaviour of the static response is
related to the second derivative with respect to the
density (or magnetization) and this provides well known
sum rules that we exploit here.
 For simplicity, and for comparison with
the degenerate two-valley case, let us review the
one-valley paramagnetic case $\zeta=0$, 
and consider the calculation\cite{cmm} of the small-$k$,
static ($\omega=0$) limit of the simple (one-valley) response function.
\begin{equation}
\chi_v(n_v)=\chi^0(n_v)/[1-V_k(1-G_{vv})\chi^0(n_v)].
\end{equation}
The density-density response function is associated with the
proper polarization function $\Pi_v$. Dropping the
species subscript $v$ for the present, we have
\begin{eqnarray}
\Pi  &=&\Pi^0/(1+V_kG\Pi^0)\\
\Pi^0&=&-\chi^0
\end{eqnarray}
The small-$k$ behaviour of this function states that
\begin{equation}
\Pi/\Pi_0=\kappa/\kappa_0 
\end{equation}
The compressibility $\kappa$ is calculated via the
chemical potential $\mu$, starting from the total free
energy per unit volume obtained from the CHNC calculation.
\begin{eqnarray}
F&=&F_0+F_x+F_c\\
F&=&\sum_v n_v[\epsilon_0^v(n_v)+\epsilon_x^v(n_v)]+n\epsilon_c(n)\\
\epsilon_0^v&=&(1+\zeta^2)/2r_{sv}^2\\
\epsilon_x^v&=&-\frac{2\surd{2}}{3\pi r_{sv}}[(1+\zeta)^{3/2}+(1-\zeta)^{3/2}]\\
\mu&=&\frac{dF}{dn}=\mu_0+\mu_x+\mu_c\\
1/\kappa&=&n^2\frac{d\mu}{dn}
\end{eqnarray}
At $T$=0, the chemical potential is given (in Hartrees) by:
\begin{equation}
\mu=n_v\pi-2(2/\pi)^{1/2}n_v^{1/2}+\mu_c
\end{equation}
The compressibility calculated directly from the 4-component calculation
should agree with that obtained from the coupled-valley formalism.
There, the small-$k$ limit of the denominator of the
density-density proper-polarization function is given by
$$1+V_kG_d\chi^0=\kappa/\kappa_0$$
Here $G_d$ is the LFF of the density-density polarization function.
Hence the denominator $d_1$, or $d_2$ 
of the $d$-$d$ response occurring in Eq.\ref{coupled},
for any particular species is available for the density-density
response function in each valley. But the cross 
density-density LFFs, e.g., $G_{12}$, and the cross-denominators
$d_{12}$ needed to form the coupled-valley forms are not yet specified.

 In the case of the spin susceptibility $\chi_s$, the role played
by the compressibility is taken over by the spin-stiffness, which
is the second derivative  of the free energy $F$ with respect to
 the spin polarization $\zeta$. Here we have, for a single 
valley,
\begin{equation}
\label{chirat}
\chi_s/\chi_P=1+d^2\{r_{sv}^2 F_{xc}\}/d\zeta^2
\end{equation}
Hence the denominators $d_u$ and $d_v$ of the spin-susceptibilities
of the  each 2DES are known,  at the valley densities 
$n_v=n_u=n/2$, from a simple CHNC calculation or from a
QMC energy parametrization. However, here again the cross terms
$d_{12}$ and $d_{21}$, (which are equal)
 needed to complete the calculation of
the coupled susceptibility (Eqs.~\ref{coupled},\ref{sumichi}) are 
not yet specified.

The cross term for the $d$-$d$ response, or for the $s$-$s$ response
can be calculated if the free-energy contribution $F_{uv}$ arising
from the Coulomb interaction among the electrons in the two
valleys is known. The interaction is among the electrons of
valley $u$, at density $n/2$, and the electrons of
valley $v$ at density $n/2$. 
The inter-valley free-energy contribution $F_{uv}(n/2,n/2)$
is purely Coulombic, and hence it is
clearly analogous to the correlation free-energy term
arising from the antiparallel-spin PDF, i.e., $g_{12}(n,\zeta=0)$ of the
simple one-valley 2DES {\it at density} $n$ with two spin species.
 The case $\zeta=0$ ensures
that the total  density is split as $n_u=n_v=n/2$. 
Thus  the $d_{12}$ term needed for the spin-susceptibility calculation
  and the
density-density response calculations are:
\begin{eqnarray}
\label{crossterm}
 s-s \;\; d_{12}&=&d^2\{r_s^2 F_c[g_{12}]\}/d\zeta^2  \\
 d-d \;\; d_{12}&=& -(2/\pi)d^2F_c[g_{12}]/dn^2
\end{eqnarray}
Note that $d_1$ is calculated from $F_{xc}(n/2,\zeta=0)$, while $d_{12}$ is
from $F_c(n,\zeta=0)$ of the simple one-valley 2DES.
Hence, knowing $d_1$, $d_{12}$, (which are equal to $d_2$, and $d_{21}$
since the valleys are degenerate), 
we can calculate the susceptibility enhancement $\chi_s/\chi_P$, as
well the compressibility ratio $\kappa/\kappa_0$ of the interacting
2-valley 2DES, without actually solving the coupled system of 10 distribution
functions needed in the full CHNC calculation of the 4-component system.
\begin{figure}
\includegraphics*[width=9.0cm, height=10.0cm]{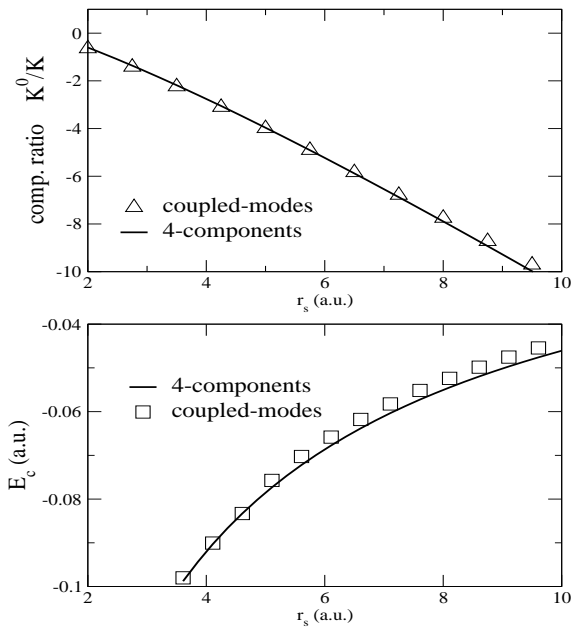}
\caption
{ (a) Comparison of the compressibility ratio $\Pi_0/\Pi_{cm}=K^0/K$ obtained
from the coupled-mode approach, and from the second-density derivative of the
total free of the 4-component system. (b) The energy estimate $\epsilon_c(rs,\zeta=0)$
from the coupled-mode form, and from the 4-component QMC\cite{cs}.
}
\label{eck0k}
\end{figure}
\subsubsection{Results for the compressibility ratio and the susceptibility ratio}
we consider the compressibility ratio $K_0/K$
obtained by the coupled-mode analysis and from the 2-valley QMC data\cite{cs}, or
equivalently, from the 4-component CHNC data, in Fig.~\ref{eck0k}.
The excellent agreement shows that the coupled-mode procedure for using the
1-valley data to generate 2-valley data is successful.

\begin{figure}
\includegraphics*[width=9.0cm, height=10.0cm]{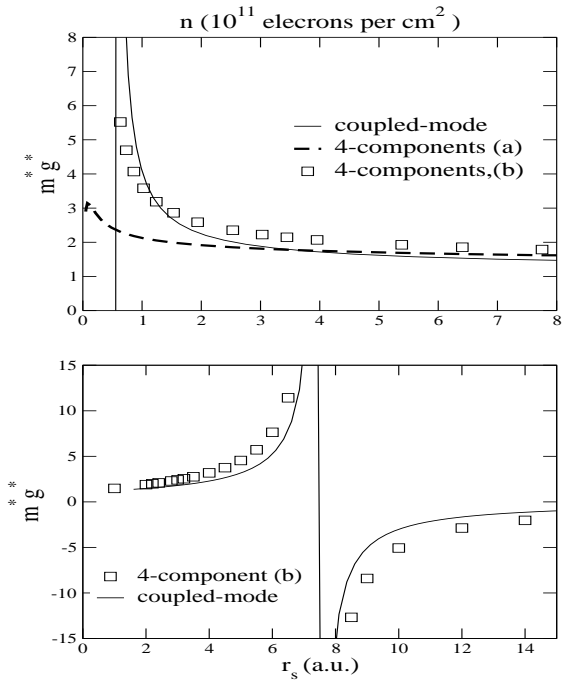}
\caption
{Comparison of the suscepribility enhancement $\chi_{cm}/\chi_P=m^*g^*$ obtained
from the coupled-mode approach and from the second-$\zeta$ derivative of the
total free of the 4-component system. The curves marked 
``4-components (a),~(b)''
 differ by $\sim$5\% in the value of the
second $\zeta$- derivative of the correlation energy,
showing the strong sensitivity to this energy derivative. 
}
\label{mgtc}
\end{figure}

The calculation of the susceptibility enhancement was given in
ref.~\cite{cmm}. We consider the susceptibility
enhancement in more detail. Equation~\ref{chirat} involves the
second $\zeta$- derivative of the correlation energy.
It is well known that the $\zeta$- dependent QMC calculations
are more prone to errors since a whole shell of spins need to be reversed.
 In fact, the 1-valley 2DES calculations of
Rapisarda and Senatore\cite{rapi} using Diffusion Monte Carlo (DMC)
predict a value of $r_s\sim 20$ for the spin transition, 
different from that ($r_s\sim 26$)
predicted by Attacalite et al.~\cite{atta} who also use very similar DMC
methods. This is an indication of the type
of uncertainty that may be had in $\zeta$-dependent QMC calculations.
 No $\zeta$-dependent QMC data are available for 
the 2-valley system.  Using Eq.~\ref{spinpol}
we can write an {\it approximate} explicit form at $T=0$:
\begin{eqnarray*}
\label{epsderiv}
\epsilon_c(r_s,\zeta)&=&\epsilon_c(r_s,0)+ 
(\epsilon_c(r_s,1)- \epsilon_c(r_s,0))p(r_s,\zeta) \\
\frac{d^2 \epsilon_c(r_s,\zeta)}{d\zeta^2}&=&\Delta E(1,0)
 \frac{d^2 p(r_s,\zeta)}{d\zeta^2}\\
\end{eqnarray*}
Thus the energy difference between the polarized phase and the  unpolarized
phase appears directly. This becomes zero in systems like the one-valley 2-DES
that show a spin transition. Even in the 2-valley system where there is
no transition to a stable $\zeta=1$ state, we can expect the calculated
$\chi_s/\chi_P$ or $\chi_{cm}/\chi_P$ to be quite sensitive to $\Delta E(1,0)$,
and to the details of the form of the $\zeta$- dependent function.
We find that this is very much the case. In Fig.~\ref{mgtc} we display
the coupled-mode evaluation of $m^*g^*=\chi_{cm}/\chi_P$ with the
value of $\chi_s/\chi_P$ obtained from the $\zeta$-second derivative of the 
energy obtained from the full 4-component calculation.
We give curves labeled ``4-component (a), and (b)'', where 
 the contribution from the $d^2\epsilon_c/d\zeta^2$- derivative
differs by $\sim$5\%. Clearly, this small change has a drastic
effect on the $m^*g^*$ evaluation.
In the lower panel of Fig.~\ref{eck0k} we compare the  2-valley correlation
energy $\epsilon_c(r_s, \zeta=0)$ from the coupled-mode analysis and from the
direct 4-component QMC calculation. As expected, a small difference appears 
at low densities, while the accuracy is good near $r_s\sim 5$
where the rapid increase in $m^*g^*$ occurs. This type of
error is quite within the
errors that are possible in the 4-component QMC or the 4-component CHNC,
and in fitting to the polarization dependence of the numerically
calculated correlation energies. Coupled-mode formation implies that the
excitation spectrum of the system no longer shows the features of the
individual valleys, and hence is consistent with the conclusions of
ref.\cite{valpud} where no evidence for intervalley scattering was seen.
\section{Relation between density $n$ and the $r_s$ parameter in $Si$-MOSFETS.}
\label{n1n2}
  In the 2DES of GaAs/AlAs- based structures, the dielectric constants of
the two materials are nearly identical. The lattice constants are also well
matched and hence the calculation of the effective atomic units needed in
converting the experimental density $n$ to the effective electron-disk radius
($r_s$) 
 can be carried out in an unambiguous, transparent way. This is
important since many-body  theory is formulated
 within the language of $r_s$ and
effective atomic units.

	Unlike in $GaAs$, the situation for $Si$-MOSFETS is more
 complicated. The
lattice mismatch between crystalline Si and most crystalline varieties
(e.g., crystobalite) of
SiO$_2$ turn out to be 35-40\%. The dielectric constants
are also strongly mismatched,
being $\sim 11.5$ and  $\sim 3.9$ for Si and SiO$_2$. The large lattice
mismatch ensures that there is {\it no} sharp Si/SiOi$_2$ interface. The 
reconstruction of the $Si$ atomic layers between the crystalline-Si and
 the SiO$_2$ bulk-like region still contain tetrahedral-bonding networks, but
with strongly modified bond angles, bond lengths etc., characteristic of the
amorphization of the $Si$ layers immediately adjacent to the oxide layer.
Many decades of experimental and theoretical work has gone into sharpening
our understanding of this interface.  More recently, 
first principles
density-functional calculations by Carrier et al.\cite{pierre}, 
starting from tight-binding models\cite{nacir}
 have presented 
a clearer picture of the atomic arrangements near the interface region.
A series of similar studies by Pasquerello et al.\cite{pasq} establish the
geometry of the Si/SiO$_2$/vacuum interface.
Thus a reliable atomic model of the Si/SiO$_2$ interface
 obtained via geometry optimization of the total energy
is now available\cite{pierre}. The essential
point is that
the Si/SiO$_2$ interface contains approximately 5 regions containing
crystalline Si (c-$Si$), amorphized Si (a-$Si$), suboxide layers,
 amorphized silicon dioxide (a-$SiO_2$),
and crystalline $SiO_2$ . These are indicated 
schematically below:
\begin{equation}
[001]\,(z\to)|\mbox{c-}Si|\mbox{a-}Si| \mbox{suboxides} 
| \mbox{a-}SiO_2 |\mbox{c-}SiO_2|
\end{equation}
The amorphous (or bond distorted) regions of a-$Si$ 
 do not have a ``conduction band'' 
and should be considered as the true insulator that
 separates the 2DES which resides
at the interface between c-$Si$ and a-$Si$. Let the
location of this amorphization edge be at $z=z_a$. This edge can be defined
to within a few  atomic planes within the first-principles theoretical models.
 If we are dealing with a thick
electron gas, then envelope-function methods for describing the form factor may
be reasonable. Otherwise a more detailed atomic description involving
Bloch functions is needed. In any case,
if the electron gas is very thin, its growth-direction
 density profile may be considered
to be $\delta(z-z_a)$. That is, crystalline $Si$ and a-$Si$ flank the
two sides of the 2-D electron layer, with  a-$Si$ playing the role of the insulator.
The valence bonds of the a-$Si$ still form a quasi-random tetrahedral network,
even though distorted, and hence the ``background'' dielectric constant of 
a-$Si$ is essentially that of $Si$ That is, the effective dielectric constant:
\begin{equation}
{\bar\epsilon}=0.5(\epsilon_{si}+\epsilon_{ins})
\end{equation}
often used for the 2DES in the MOSFET positioned at
 an abrupt $Si/SiO_2$ interface
should be reconsidered. The second formula of Ando et al.
(see appendix of ref.\cite{afs}) for the conversion $n$ to $r_s$,
 using a mean ${\bar\epsilon}$ of 7.7 
 is not recommended. Instead, the
first formula of ref.\cite{afs}, i.e.,
\begin{equation}
\label{firstformula}
r_s/a^*=1.751\left[\frac{10^{12}\mbox{cm}^{-2}}{n_s}\right]^{1/2}
\left[\frac{11.5}{\epsilon_{sc}}\right] 
\left[\frac{m}{0.19m_0}\right]
\end{equation}
is clearly the one consistent with the first-principles
atomic structure of the $Si/SiO_2$ interface referred to above\cite{pierre}.
If we look at the $Si$-MOSFET literature, we find that the formula
which uses the average dielectric constant of 7.7, appropriate for the abrupt
$Si/SiO_2$ interface has been used by a number of authors. These authors use
a value of $r_s$ increased by a factor of $\sim 1.49$ compared to
what we recommend.
 Thus Pudalov et al.\cite{pudalov}, and Okamoto et al.\cite{okamoto} have used the
mean dielectric constant of 7.7 for their calculation of $r_s$. However, both
these studies use the $r_s$ parameter mainly as a plotting variable in the
figures, and not for
any many-body calculations. Hence a choice of $r_s$
 which differs from that used in our work by a factor of $\sim$1.49
is immaterial. 
In the review article by Kravchenko and Sarachik\cite{krav}, 
values of $r_s$ are further
modified by the experimentally obtained $m^*$ to
 discuss the interactions in $Si$ MOSFETS.
Thus their  $r_s^*$ is not used in the sense used in standard many-body theory.
 Hawang and Das Sarma\cite{dasrs} have also examined $Si$-MOSFET
 resistivities, using
an impurity-scattering calculation which requires defining
 the effective background
dielectric constant. They point out that their results are qualitative.  
Their results
would not be affected by the choice of either formula given by ref.\cite{afs}, 
i.e, using ${\bar\epsilon}$ =7.7 or 11.5. Altshuler and Maslov\cite{maslov}
actually consider the implications of the suboxide layer and how this could play a
significant role in the theory. However, they too point out that their effort
 is essentially  to
indicate  a ``mechanism'' rather than a theory of
 the metal-insulator physics of $Si$-MOSFETS.
Hence, once again, the results are too qualitative to make any difference.
Similarly,
the results of other workers \cite{punnose} also do not discriminate sufficiently
to make the choice of ${\bar\epsilon}$ a significant issue. 

Another class of problems where the choice of the average dielectric
 constant is an issue
is in calculating electric subband energies\cite{afs}. Although the eigenvalues
 of the Kohn-Sham
equation are not to be considered as effective excitation energies,
such an assumption 
is often made. The input dielectric constant
 (which decides the effective $r_s$) enters into
the exchange-correlation functions as well as the Poisson potential
 used. Most calculations
are in the high density regime, and  it turns out that, 
given the uncertainties
of the quantum-well potentials and other parameters,
 the  results can be equally well explained by
a range of values of the effective dielectric constant. 

This situation becomes quite different when it comes to  {\it quantitative}
 calculations for {\it low-density}
MOSFETs, e.g., in the regime $n=1\times10^{11}$ electrons/cm$^{2}$. Our CHNC calculations
for $m^*$ and $g^*$ presented in ref.\cite{cmm} clearly favour the first formula,
Eq.~\ref{firstformula} of
Ando et al.\cite{afs} as the correct formula. 
This is also the formula that is consistent
with the Car-Parrinello optimized atomic structure
 of the $Si/SiO_2$ interface obtained
from the calculations of Carrier et al.\cite{pierre}
 and  also of Pasquerello et al.\cite{pasq}).
 We believe that the problem of the correct dielectric constant at the
Si/SiO$_2$ interface has received little scrutiny within the
2-D electron community in the past because there
was no analytic many-body theory  capable
 of giving quantitative results for low-density electron
systems.  Also, it is interesting to note that if the second formula of Ando et al. were
used instead of the first formula that we recommend, then the calculated  {\it total}
 $r_s$ is close to the $n/2$ value (per valley, $\sim 1.414r_s$) of $r_s$
 calculated by the first formula.  This
is consistent with the calculations that we advocate, at the Hartee-Fock level (i.e,
at the single-electron level). This fact can also lead to some confusion in assessing
the validity of numerical calculations.

The CHNC programs for electron gas calculations mentioned here and in
ref.~\cite{prl2} may be accessed via the
internet at the address given in Ref.~\cite{web}.

\section{Conclusion}
 We can derive the following conclusions from this study. The CHNC method applied
to a four-component electron fluid gives results in very close agreement with
available Diffusion Monte Carlo calculations,
without the use of any adjustable parameters
specific to the 2-valley problem. The coupled-mode approach to constructing the
2-valley properties from 1-valley data is also fully confirmed. The calculation of the
spin-susceptibility enhancement $\chi_s/\chi_P$ from the
 second $\zeta$- derivative of the spin
dependent energy is found to be very sensitive to
 the energy difference between
the polarized and unpolarized phases and to the form of the
polarization dependence. Nevertheless, the coupled-mode form is very
successful in capturing the required physics. Thus the QMC, the 2-valley CHNC, 
and the coupled-mode approach based on the 1-valley data provide
three independent methods for the study of the strongly
coupled 2DES in Si MOSFETS. The three method are found to
be in excellent agreement. Finally, we note that the methods
used in this paper can be used to study electron/hole bilayers
which are separated by a physical distance $d_L$, the present work being
for $d_L=0$.

\end{document}